\documentstyle[twoside,fleqn,espcrc2,fps]{article}

\newcommand{\Z}{Z \!\!\! Z}

\newcommand{\AmS}{{\protect\the\textfont2
  A\kern-.1667em\lower.5ex\hbox{M}\kern-.125emS}}

\title{Perfect Actions for Scalar Theories
        \thanks{Talk presented at LAT97.}}

\author{W. Bietenholz, HLRZ c/o Forschungszentrum J\"{u}lich,
52425 J\"{u}lich, Germany}

\begin{document}

\begin{abstract}
We construct an optimally local perfect lattice action for free
scalars of arbitrary mass, and truncate its couplings to a unit
hypercube. Spectral and thermodynamic properties of this ``hypercube
scalar'' are drastically improved compared to the standard action.
%Preconditioning can still be applied to this action, which has a 
%fermionic analog. 
We also discuss new variants of perfect actions,
using anisotropic or triangular lattices, or applying
new types of RGTs. 
%motivated by the B-spline formalism.
Finally we add a $\lambda \phi^{4}$ term
and address perfect lattice perturbation theory.
% In particular
We report on a lattice action for the anharmonic oscillator, which
is perfect to $O(\lambda )$. 
%It involves 2- and 4-variable couplings
%over maximally two lattice spacings.

\end{abstract}

\maketitle

Many examples have shown that the perfect action program to
construct improved lattice actions works beautifully, if it 
can be properly implemented. However, a convincing 
application to QCD is still outstanding. Such attempts
are a desperate struggle for locality and questions of
parameterization and truncation are major issues.
Hence it is motivated to study the properties
of perfect action carefully in simple situations.

For free and perturbatively interacting fields, perfect
action can be constructed conveniently by a technique that
we call ``blocking from the continuum'' \cite{QuaGlu}. 
It corresponds to a
blocking factor $n$ RGT in the limit $n\to \infty$, so that
the RGT does not need to be iterated.
For a scalar we can relate
the continuum resp. lattice field $\varphi, \ \phi$ as
\vspace*{-2mm}
\begin{equation}
\phi_{x} \sim \int \Pi (x-y) \varphi (y) dy \ ,
\end{equation}
\vspace*{-2mm}
where $x \in \Z^{d}$ and 
$\Pi (u) \doteq \prod_{\mu =1}^{d} \Theta (1/2- \vert u_{\mu}\vert )$.
%$\Pi (u) = 1$ if $\vert u_{\mu} \vert
%\leq 1/2, \ \mu =1 \dots d$ and 0 otherwise.

If we implement this relation by a Gaussian RGT term 
-- with coefficient $1/\alpha$ --
then we obtain in  momentum space the perfect lattice
action
\vspace*{-2mm}
\begin{eqnarray}
S [ \phi ] &=& \frac{1}{(2\pi )^{d}} \int_{B} dk \frac{1}{2}
\phi (-k) G^{-1}(k) \phi (k) \ , \nonumber \\
G(k) &=& \sum_{l\in \Z^{d}} \frac{\Pi^{2}(k+2\pi l)}{(k+2\pi l)^{2}
+m^{2}} + \alpha \label{perfact} \ ,
\vspace*{-4mm}
\end{eqnarray}
%\vspace*{-2mm}
where $\Pi (k) = \prod_{\mu} \hat k_{\mu}/k_{\mu}$,
$\hat k_{\mu} = 2 \sin (k_{\mu}/2)$ and
$B = ]-\pi ,\pi ]^{d}$.
Remarkably, the spectrum of this action,
$E^{2}(\vec k ) = (\vec k + 2\pi \vec l )^{2}+m^{2}$,
is exactly the continuum spectrum, which shows that the
perfect action can reproduce the full Poincar\'{e} symmetry
in observables, even though this symmetry is not manifest
in the action.

\begin{figure}[hbt]
\vspace{-8mm}
\def\fpsangle{0}
\epsfxsize=70mm
\fpsbox{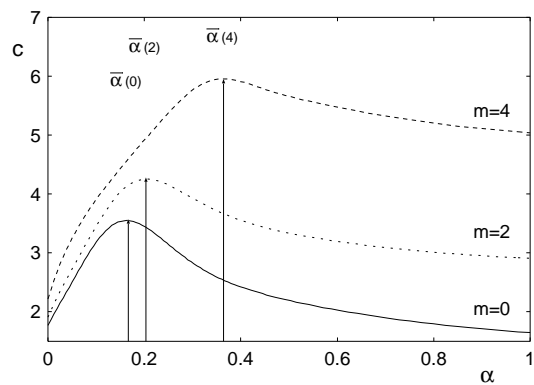}
\vspace{-10mm}
\caption{The locality parameter $c (\alpha )$.}
\label{smdec}
\vspace{-8mm}
\end{figure}

If we choose the RGT parameter $\alpha = 
\bar \alpha (m) \doteq (\sinh (m) -m)/m^{3}$,
then the couplings in
\vspace*{-2mm}
\begin{equation}
S[\phi ] = \frac{1}{2} \sum_{x,y} \phi_{x} \rho (x-y) \phi_{y}
\end{equation}
\vspace*{-2mm}
are restricted to nearest neighbors in $d=1$ \cite{AT}, 
and they decay exponentially and very fast in $d>1$.
\footnote{For a finite blocking factor $n$ and $m=0$, 
$\bar \alpha$ is 
replaced by $\bar \alpha_{n} 
%= \bar \alpha (1 - 1/n^{2})
= (1-1/n^{2})/6 $, in agreement with \cite{BeWi}.}
The question arises, if this choice is really {\em optimal}
for locality in $d=4$, as we postulated before for fermions
\cite{QuaGlu}.
We measure the decay by $\rho (i,0,0,0) \propto \exp \{ 
-c(\alpha )i \}$, and Fig. \ref{smdec} shows $c(\alpha )$ for various
masses. Indeed, the peaks are just at $\bar \alpha (m)$.
\footnote{For heavy fermions, 
on the other hand, we noticed that locality 
can be improved slightly beyond the 1d formula.}

\begin{figure}[hbt]
%\vspace{-8mm}
\def\fpsangle{0}
\epsfxsize=70mm
\fpsbox{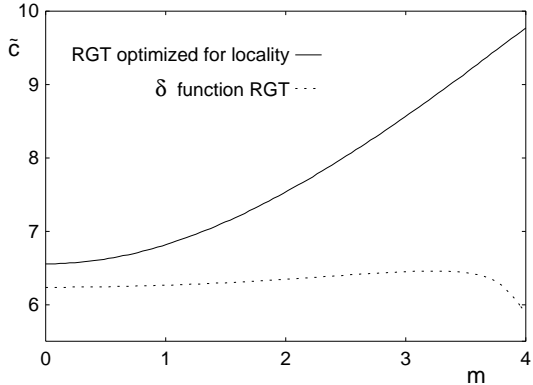}
\vspace{-10mm}
\caption{The locality measured by $\tilde c (m)$ in
$\rho (i,i,i,i) \propto \exp (-\tilde c (m)i)$, for $\bar \alpha (m)$
and for $\alpha = 0$.}
\label{decm}
\vspace{-8mm}
\end{figure}

As in the case of fermions, locality becomes even better for
increasing mass, if we use $\bar \alpha (m)$, see Fig. \ref{decm}.
 
To make this action applicable we need to truncate the couplings
to a short range. We do so by imposing periodic boundary conditions
over 3 lattice spacings, and use the resulting couplings then
in any volume \cite{LAT96}. In contrast to a truncation in c-space,
$\rho (r)$ vanishes smoothly at the edges of a unit hypercube,
the normalization is automati-cally correct and the ``decimation in
p-space'' is handy in perfect lattice perturbation theory.

The resulting couplings represent ``smoothly smeared'' lattice
derivatives. Alternatively there are various ways to apply Symanzik's
program to cancel the standard $O(a^{2})$ artifacts. 
Symanzik himself suggested
to use additional couplings at distance 2 on the axes.
These couplings are completely different from the perfect truncated
ones. If, however, one constructs a Symanzik improved hypercube
scalar, then the couplings look very similar \cite{WB}.
The same behavior
has been observed for staggered fermions \cite{FK}.
Other Symanzik improved fermions in the literature
(D234, Naik) are constructed along the axes again. 
However, I would rather
recommend to use couplings in the unit hypercube,
which is e.g. more promising
for the restoration of rotational invariance.

\begin{figure}[hbt]
%\vspace{-8mm}
\def\fpsangle{0}
\epsfxsize=70mm
\fpsbox{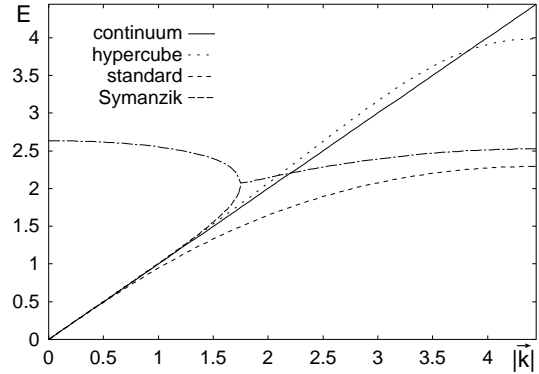}
\vspace{-10mm}
\caption{The spectrum along the (110) direction at $m=0$.}
\label{spec0}
\vspace{-8mm}
\end{figure}

In Fig. \ref{spec0} %and \ref{spec2} 
we compare the {\em dispersion relations} %at $m=0$
of our hypercube scalar, the standard formulation
and the Symanzik scalar on the axes. As in the fermionic case 
\cite{LAT96}, the latter is good at small momenta, until it is hit by
a ``ghost'' and then the solutions become complex, i.e. useless.
The hypercube scalar, on the other hand, behaves very well
all the way up to the edge of the Brillouin zone.
This behavior persists at $m>0$.

%\begin{figure}[hbt]
%%\vspace{-6mm}
%\def\fpsangle{270}
%\epsfxsize=50mm
%\fpsbox{dispm2a.eps}
%\vspace{-10mm}
%\caption{The spectrum along the (110) direction at $m=2$.}
%\label{spec2}
%%\vspace{-8mm}
%\end{figure}

\begin{figure}[hbt]
\vspace{-8mm}
\def\fpsangle{0}
\epsfxsize=70mm
\fpsbox{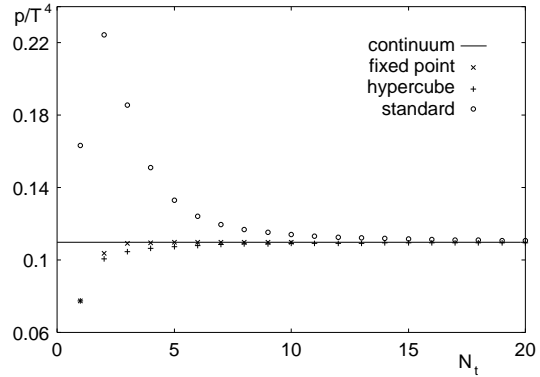}
\vspace{-10mm}
\caption{The ratio $p/T^{4}$ depending on $N_{t}$. }
\label{SB}
\vspace{-8mm}
\end{figure}

The hypercube scalar has also good {\em thermodynamic} properties.
Fig. \ref{SB} shows the ratio
pressure/(temperature)$^{4}$, which is $\pi^{2}/90$ in the continuum.
For a small number $N_{t}$ of discrete points in Euclidean
time, this ratio is approximated well for the hypercube scalar,
whereas the standard action requires a large $N_{t}$ to converge.
Other quantities like the energy density look similar
\cite{WB}.

In thermodynamics it is fashionable to use {\em anisotropic}
lattices. There is no additional problem to put a perfect
action on an anisotropic lattice. In the most general case,
lattice spacings $(a_{1},\dots ,a_{d})$,
we just have to substitute in eq. (\ref{perfact}) $l \to
(l_{1}/a_{1}, \dots , l_{d}/a_{d})$, $\hat k_{\mu} \to
(2/a_{\mu})\sin (k_{\mu}a_{\mu}/2)$, and with $\bar \alpha \to
\bar \alpha /a_{\nu}^{2}$ the mapping on the $\nu$ axis is ultralocal.
In the typical case one has $a_{d} < a_{s} \equiv a_{spatial}$, and one
might be afraid of time-like ``ghosts''. One can avoid them
by giving up perfection in the temporal direction,
\vspace*{-2mm}
\begin{displaymath}
G_{ani}(k) \! = \!\!\! 
\sum_{\vec l \in \Z^{d-1}} \! \frac{\prod_{\nu =1}^{d-1}
\hat k_{\nu}^{2}/(k_{\nu}+2\pi l_{\nu} /a_{s})^{2}}
{(\vec k + 2\pi \vec l /a_{s})^{2}+ \hat k_{d}^{2}+m^{2}}
+ \frac{\alpha}{a_{s}^{2}} .
\end{displaymath}
\vspace*{-1mm}

We can also put the perfect action on a {\em triangular} lattice.
In $d=2$ we obtain a modified $\Pi$ function,
\ \ $\Pi_{tria}(k) \ = $
%\begin{eqnarray*}
%&& \hspace{-7mm} \Pi_{tria}(k) = \\
%&& \hspace{-7mm} 
\vspace*{-2mm}
\begin{displaymath}
\frac{8[
k_{1}\cos\frac{k_{1}}{2} + k_{2} \cos \frac{k_{2}}{2}
- (k_{1}+k_{2}) \cos \frac{k_{1}+k_{2}}{2}]}
{3k_{1}k_{2}(k_{1}+k_{2})} ,
\end{displaymath}
\vspace*{-2mm}
%\end{eqnarray*}
where we refer to axes crossing under $\pi /3$.
The blocking from the continuum then leads to
\vspace*{-2mm}
\begin{eqnarray*}
\phi (k) & \sim & \frac{3}{4} \sum_{l \in \Z^{2}}
\varphi (k_{l}) \Pi_{tria}(k_{l}) \\
k_{l} &\doteq & k + \frac{4\pi}{3}
\left( \begin{array}{cc} 2 & -1 \\ -1 & 2 \end{array} \right) l \\
G_{tria}(k) &=& \!\!\!\! \sum_{l \in \Z^{2}}
\frac{\Pi_{tria}^{2}(k_{l})}{k_{l,1}^{2}+k_{l,2}^{2}+
k_{l,1}k_{l,2}+m^{2}} + \alpha .
\end{eqnarray*}
%\vspace*{-1mm}
Again the spectrum is perfect; we have e.g. full
rotational invariance in the observables, although the
lattice structure is visible in the action.

Back to the hypercubic lattice with spacing 1: we can also vary
the convolution function in the blocking from the continuum,
\vspace*{-2mm}
\begin{displaymath}
\phi_{x} = \int dy \Big[ \prod_{\mu} f_{n}(x_{\mu}-y_{\mu}) \Big]
\varphi (y) .
\end{displaymath}
\vspace*{-2mm}
Sensible candidates are the B-spline functions $f_{0}(s)
= \delta (s)$, $f_{n+1}(s) = \int_{s-1/2}^{s+1/2}f_{n}(t)dt$.
The appropriate convolution function for the gauge field $A_{\mu}$
then reads $f_{n+1}(u_{\mu}) \prod_{\nu \neq \mu} f_{n}(u_{\nu})$.
This set of functions has the correct normalization and a ``democracy
of the continuum points'', which all contribute with the
same weight to the lattice variables. In the propagator
of eq. (\ref{perfact}), the power of the $\Pi$ function generalizes
to $2n$. Deci-mation ($n=0$) fails because the sum over $l$
diverges in $d>1$, and $n=1$ is what we had before. For higher $n$
we can still achieve ultralocality in $d=1$ by adding kinetic
terms in $\alpha$, but the locality in $d>1$ is best for $n=1$.
Generally, overlapping blocks seem to be unfavorable for locality.
We can contract $f_{n}$ into [-1/2,1/2], and re-adjust the 
normalization (giving up ``democracy''). Thus $f_{2}$ turns
into an ``Eiffel tower function''
$\tilde f_{2}(s) = (2- 4 \vert s \vert) \Theta(1/2-\vert s \vert )$, 
which is in business
in view of the induced locality \cite{WB}. For $n>2$ one obtains
blocking schemes which are difficult to relate to finite
blocking factors, hence they are not easily compatible 
with the multigrid improvement. So I recommend
the ``Eiffel tower function'' $\tilde f_{2}$ as a promising
alternative to $f_{1}$.\\

\vspace*{-3mm}
Finally we proceed to the $\lambda \phi^{4}$ theory.
If we block perturbatively from the continuum, we
encounter divergent loop integrals, which can be
regularized 
%by a standard technique 
in the continuum.
There is no divergence, however, for the {\em anharmonic
oscillator} ($d=1$). There we constructed an $O(\lambda )$
perfect action. It involves 2- and 4-variable couplings
$\propto \lambda$, which have an analytic form in momentum space
and which we evaluated numerically in c-space.
For $\bar \alpha (m)$ they do not couple any variables
over distance $>2$ (the perfect action to $O(\lambda^{n})$
extends to maximal distances $2n$ ($n\geq 1$)).
 
The performance of this action is of interest in view
of the direct application of the perfect quark-gluon vertex 
function in QCD \cite{KO}.
We measured the first two energy gaps \cite{thor},
$\Delta E_{1}$ and $\Delta E_{2}$, for our action
and for the standard action, and observed that they
are clearly closer to the continuum results for the
new action up to $\tilde \lambda \doteq \lambda /m^{3}
\simeq 0.25$. This corresponds to an improved
asymptotic scaling.
However, for the scaling quantity $\Delta E_{2}/
\Delta E_{1}$ the improvement is difficult to demonstrate,
because it is restricted to tiny $\tilde \lambda$. 
%(and even
%there the standard action is also quite good).
In both cases, the $O(\tilde \lambda )$ perfect action
is successful up to the magnitude of $\tilde \lambda$,
where also first order continuum perturbation theory -- for the
considered observable -- collapses.
%Both of these observations agree
%with continuum perturbation theory, i.e. the $O(\lambda )$
%perfect action works -- for a certain observable --
%up to the magnitude of $\tilde \lambda$,
%where $O(\lambda )$ continuum perturbation theory 
%for that observable collapses.
%just as successful as continuum perturbation theory
%to $O(\lambda )$.

\end{document}